# RF-driven modification of phase space distribution

A. Burov, Fermilab[1]

**Abstract**

It is shown that bunch phase space distribution can be modified by means of RF phase modulation.

## 1. Motivation and method

Tevatron bunches are known to demonstrate longitudinal instability sometimes called as "dancing bunches" [1]. There are some theoretical indications showing that this instability is sensitive to a derivative of the phase space density $f'(J)=df/dJ$ at small actions $J$, i.e. $f'(J\to 0)$. Distributions with flat or even positive slope of the distribution density at small actions appear to be more beneficial [2]. However, intra-beam scattering always tries to make this derivative negative, $df/dJ \cong f/J_{rms}$. Thus, the problem emerges - is it possible to zero this derivative or even change its sign by one or another means?

It is known that the distribution function tends to flatten inside resonance separatrices. This leads to an idea, that desired change of the distribution may be achieved by introducing a proper resonance for small amplitude particles. Due to nonlinearity of the synchrotron motion, large amplitude particles could be only slightly disturbed by that. Technically, easiest way for that would be modulation of the RF phase with a synchrotron frequency for small amplitude particles. Then, a width of the affected area would be determined by amplitude of this phase modulation $\varphi_0$.

For this sort of perturbation, the following mapping applies:

$$
\begin{aligned}
z_{n+1} &= z_n + \Delta t \cdot p_n \\
p_{n+1} &= p_n - \Delta t \cdot \sin\left(z_{n+1} - \varphi_0 \sin(t_n)\right) \\
t_{n+1} &= t_n + \Delta t
\end{aligned}
\qquad (1.1)
$$

Here time $t$ is measured in those units that the small-amplitude synchrotron angular frequency $\omega_s = \nu_s \omega_0 = 2\pi\nu_s / \Delta t = 1$. In other words, in these units the revolution time $\Delta t$ is equal to the synchrotron phase advance per revolution, $\Delta t = 2\pi\nu_s$. As it is seen from the RF term in that mapping, the coordinate z is measured in units of RF phase, $-\pi < z < \pi$.

---

## 2. Results

To see how this mapping changes the distribution function, tracking of N=$10^4$ particles was simulated on a base of *Mathematica*. Original distribution over unperturbed actions $J$ was taken as $f(J) \propto \left(1 - J/J_{lim}\right)^2$, $J \leq J_{lim}$, with the total emittance $J_{lim} = 2$, close to the bucket acceptance $J_{max} = 8/\pi \approx 2.55$. To see an example seemingly close to desired strength of the perturbation, the amplitude of the RF phase modulation was taken $\varphi_0 = 0.05$. The synchrotron phase advance was taken as $\Delta t = 0.025$, and the simulation time Tsim=500 synchrotron radians =80 synchrotron periods. Histograms for initial and final action and phase probability distribution functions (PDF) are shown in Figs. 1 and 2.

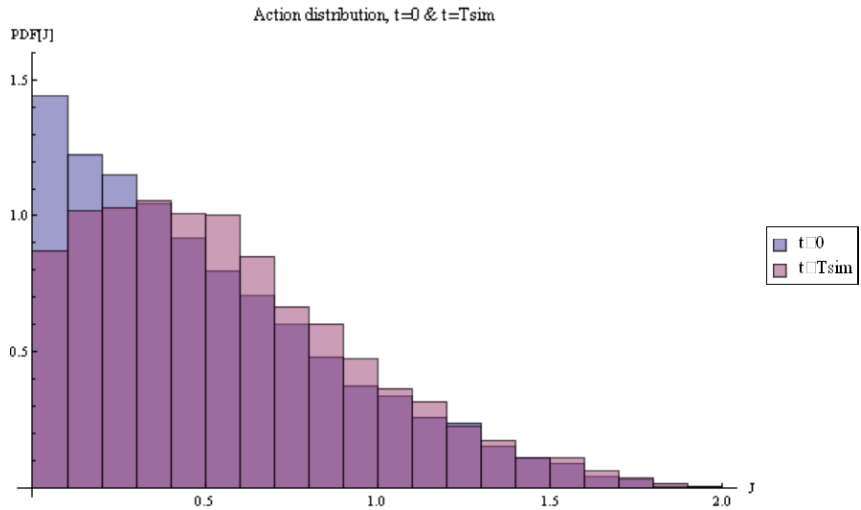

Fig. 1. Initial (blue) and final (pink) distributions over action. Overlapping area is violet.

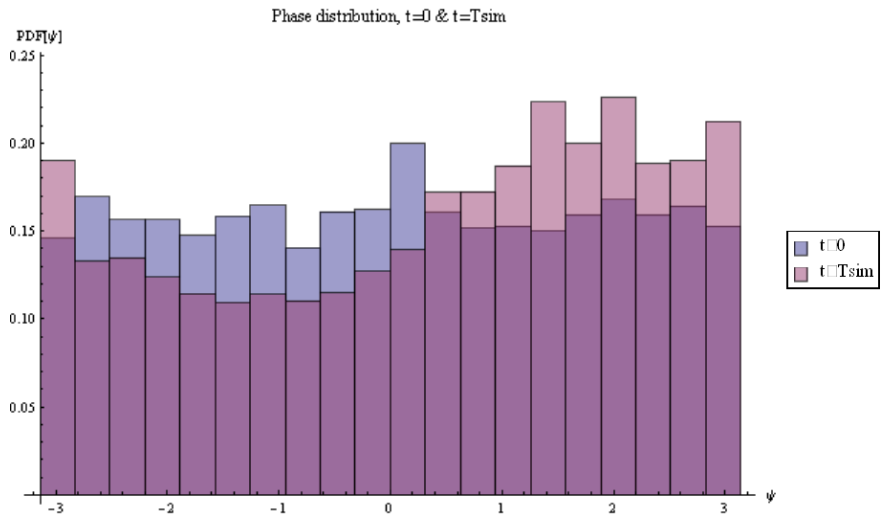

Fig. 2. Initial (blue) and final (pink) distributions over phase.

These two figures show a couple of interesting things. First of all, the final action distribution at small actions is either flattened or may even change a sign of its derivative – so the original goal is reached. Second, the final phase distribution is less even than the original one; there is some coherent dipole motion in the final state. Time evolution of this coherent motion is presented in Fig. 3 as a plot for the centroid Hamiltonian

$$H(<z>,<p>) = \frac{<p>^2}{2} + 1 - \cos(<z>),$$

where $<z>$ and $<p>$ are ensemble-averaged instantaneous values for the offset $z$ and momentum $p$.

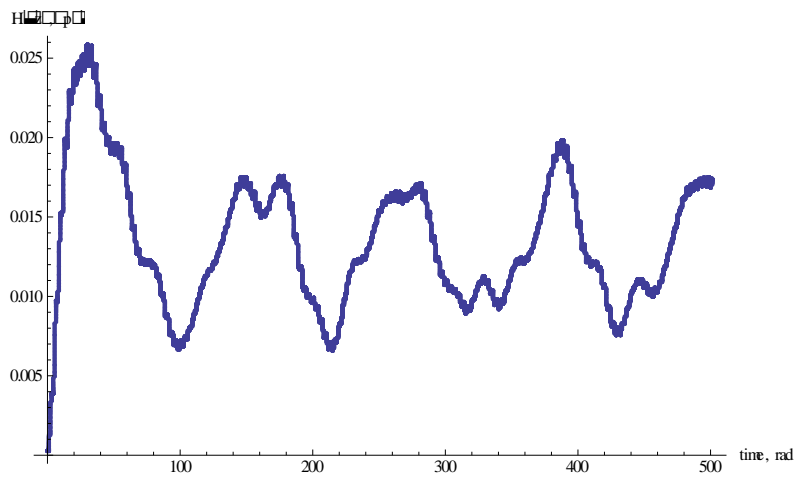

Fig. 3. Centroid Hamiltonian as a function of time. Time average $\bar{H}(<z>,<p>) \approx 0.01 H_{lim}$, where $H_{lim} = 1.7$ is the Hamiltonian at the distribution border $J_{lim} = 2$.

Figures 4 and 5 give a comparison of the final action and phase distribution with its transformation after a quarter of the synchrotron period. As it can be expected, the two action distributions are almost identical, while the two phase distributions show a shift of the phase wave by about π/2.

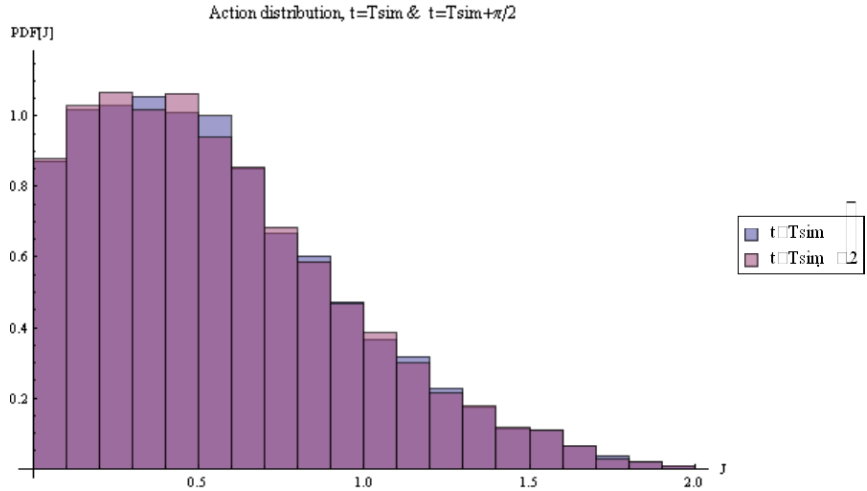

Fig. 4. Final action distribution (blue) and its transformation after a quarter of the synchrotron period (pink). The two distributions are almost identical, as expected.

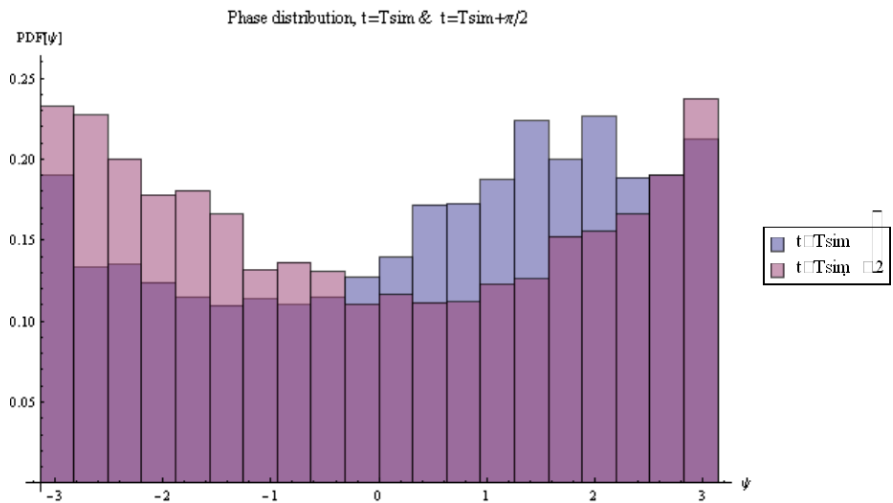

Fig. 5. Same for the two phase distributions. There is an expected phase shift.

## 3. Discussion

RF phase modulation is clearly a working tool for flattening of the distribution function. From this point of view, it should help to stabilize longitudinal instability [2]. Although RF phase modulation itself excites coherent motion at some level, it should not make a significant problem. Indeed, this RF modulation is needed only for ~ 100 synchrotron revolutions or so, and after that it has to be switched off. The provided distribution should be stable, so any coherent motion should decay.

The applied mapping Eq. (1.1) does not take into account the ring impedance. Impedance leads to potential well distortion, depressing synchrotron frequencies. This circumstance slightly reduces the required frequency of the RF modulation.

Dedicated beam studies would be able to shed more light on this issue.

## References

[1] R. Moore et al., "Longitudinal bunch dynamics in the Tevatron", Proc PAC 2003, p. 1751.
[2] A. Burov, to be published.